\documentclass[12pt]{article}
\pdfoutput=1
\usepackage{graphicx} 
\usepackage{amsmath,amsfonts,amssymb,stmaryrd,latexsym}
\usepackage{url,subcaption,comment}
\usepackage{hyperref} \usepackage{mathtools} \usepackage{xcolor} 
\hypersetup{
    colorlinks=true,       % false: boxed links; true: colored links
    linkcolor=purple,    %red,  %  magenta,      %violet,          %   %black,          % color of internal links
    citecolor=blue,        % color of links to bibliography
    filecolor=magenta,      % color of file links
    urlcolor=black           % color of external links
}
\usepackage{cite}
\usepackage{dcolumn}% Align table columns on decimal point
\newcommand{\be}{\begin{equation}}
\newcommand{\ee}{\end{equation}}
\newcommand{\bear}{\begin{eqnarray}}
\newcommand{\eear}{\end{eqnarray}}
\newcommand{\ba}{\begin{array}}
\newcommand{\ea}{\end{array}}

\begin{document}

\baselineskip=18pt \setcounter{page}{1}

\vspace*{-1.2cm}

\noindent \makebox[11.9cm][l]{\small \hspace*{-.2cm} }{\small  
Fermilab-Pub-21-771-T}  \\  [-1mm]

\begin{center}

{\Large \bf  
A model of quark and lepton compositeness
}    \\ [9mm]

{\normalsize  \bf Bogdan A. Dobrescu  \\ [3mm]
{\small \it Particle Theory Department, Fermilab, Batavia, IL 60510, USA     }}\\

\vspace*{0.8cm}

{\normalsize  December 30, 2021 }

\end{center}

 \vspace{0.1cm}

\renewcommand{\abstractname}{\vspace{-\baselineskip}}
\begin{abstract} \normalsize
In the chiral $SU(15)$ gauge theory presented here, the quarks and leptons are bound states (``prebaryons") of massless preons.  
The Standard Model charges of the preons imply 3 generations of quarks and leptons, plus some vectorlike fermions
 lighter than the confining scale $\Lambda_{\rm pre}$.  Under certain assumptions about the chiral dynamics, 
 bound states of two prebaryons behave as Higgs fields.  The  QCD and electroweak
groups may unify above $\Lambda_{\rm pre}$, while $SU(15)$ prevents rapid proton decay.
\end{abstract}
 
\medskip

%%%%%%%%%%%%%%%%%%%%%%%%%%%%%%%%%%%%%%%%%%%%%%%%%%%%%%%%%%%%%
\section*{\large 1. \ Composite chiral fermions}  

\vspace*{-2mm}  
  
Since the nucleons are made of quarks, 
it is compelling to ask whether quarks might have substructure.  If they do, then leptons should  also 
be composite particles, given the correlations  between quark and lepton charges under the Standard Model (SM) gauge group,
$SU(3)_c\times SU(2)_W\times U(1)_Y$. Developing a self-consistent theory of quark and lepton compositeness is, however, a daunting task.
An obstacle is the chiral nature of the SM fermions, which implies that the dynamics responsible for their compositeness must also be chiral. Unfortunately, the behavior of strongly-coupled chiral theories remains uncertain.  
Nevertheless, some consistency checks of the possible chiral dynamics have been devised \cite{tHooft:1979rat, Dimopoulos:1980hn, Eichten:1985fs}, 
pointing to the probable spectrum of light  bound states. 

Even so, it remains difficult to see how the peculiar pattern of quark and lepton masses could arise from 
compositeness. If the quarks and leptons 
are made of massless fermions (traditionally called ``preons"\cite{Pati:1975md}) as expected in chiral 
gauge theories, then the quarks and leptons are likely to remain massless, or at least degenerate in mass.

A model of quark and lepton compositeness, with several realistic features, is presented here. 
The strong coupling dynamics is based on a preonic $SU(15)$ gauge group.
The global symmetry of the preons has room for the 
quantum numbers of a single generation of fermions, but the chiral preonic baryons (``prebaryons") include all 3 SM generations.
Higgs doublets arise as bound states of two prebaryons,  reminiscent to the formation of deuteron within QCD. 
In this model,  QCD loses asymptotic freedom near the $SU(15)$  confining  scale, $\Lambda_{\rm pre}$. 
However,  the SM gauge groups may unify at a scale above $\Lambda_{\rm pre}$, recovering asymptotic freedom. 
Interestingly, the unification scale can be lower than usual because the $SU(15)$ 
symmetry protects against rapid proton decay. 

\medskip
    
%%%%%%%%%%%%%%%%%%%%%%%%%%%%%%%%%%%%%%%%%%%%%%%%%%%%%%%%%%%%%%
\section*{\large 2. \ Preonic $SU(15)$ gauge dynamics}
%\label{sec:pre}

The preonic gauge group  must be asymptotically free, so that the preons may confine, and the fermion representations must be free of gauge anomalies.
It has been observed \cite{Dimopoulos:1980hn, Eichten:1985fs, Geng:1986xh} that $SU(N)$ gauge theories with  
a fermion in the symmetric representation 
and $N+4$ fermions in the fundamental representation likely produce massless chiral baryons. 
Other gauge groups or preon representations are possible \cite{Bars:1981se, Georgi:1985hf}, but the dynamics is less certain.  

The model proposed here has one left-handed fermion, $\Omega$,  in the symmetric 2-tensor conjugate representation of the preonic $SU(15)_{\rm p}$ gauge group. That representation has dimension 120, and its anomaly 
is cancelled by 19 left-handed fermions, $\psi_i$,  in the fundamental representation. Thus, the global symmetry, 
$SU(19) \times U(1)$, is large enough to embed   the SM gauge group, as shown in Table \ref{table:preons}.
The $\psi_i$ preons with $i = 5, ..., 19$ are relabelled to indicate that they carry the charges of a SM generation.

%%%%%%%%%%%
\begin{table}[b!]
\begin{center}
\renewcommand{\arraystretch}{1.5}
\begin{tabular}{|c|c|c|c|}\hline  
 Fermion    &   $SU(15)_{\rm p}$   &   $SU(3)_c \times SU(2)_W $   &  \  $U(1)_Y$    \
\\ \hline \hline
    $ \psi_Q $   &  15   &   $(3 , 2)$  &   $+1/6$
 \\ \hline 
    $ \psi_U $   &  15   &   $(\overline  3 , 1) $ &    $-2/3$
 \\ \hline 
    $ \psi_D $   &   15   &  $( \overline 3 , 1) $  &      $+1/3$
 \\ \hline 
    $ \psi_L$  &   15   &  $( 1 , 2) $  &        $-1/2$
 \\ \hline 
    $ \psi_E$  &   15   &  $( 1 , 1) $  &   $+1$ 
 \\ \hline 
   \ $ \psi_1,  ... ,   \psi_4     $ \
     &   15   &  $( 1 , 1) $  &   0
 \\ \hline 
  $\Omega$  &    $\overline{120}$    &   $(1 , 1)$  &  0     
\\   \hline
\end{tabular}
% \vspace*{-0.1cm}
\caption{Preons charged under the confining gauge group $SU(15)_{\rm p}$, and their SM charges. \\[-0.7cm] }
\label{table:preons}
\end{center}
\end{table}

Assuming that the $SU(15)_{\rm p}$ interactions are confining, the bound states made of three preons are 
$\psi_i \psi_j \, \Omega $, with $i\neq j$ and $i,j = 1, ... 19$.
These belong to the 171-dimensional antisymmetric  representation of global $SU(19)$. 
A key test that the 171 prebaryons remain massless when the preons confine
is the 't Hooft anomaly matching \cite{tHooft:1979rat}, {\it i.e.}, the condition that the global anomalies of the preons 
are equal to those of the prebaryons. Consider the global 
$U(1)$ under which all $\psi_i$ preons have charge $z_\psi$, and $\Omega$ has charge $z_{_\Omega} $.
The $[U(1)]^3$ anomalies of the preons, and of the  prebaryons are  \\ [-4mm]
\bear   
A_{\rm preon} =  285   z_\psi^3 +   120  z_{_\Omega} ^3   ~~,
\nonumber \\ [-2.mm]
\\ [-2.5mm]
A_{ \psi\psi \Omega} = 171  \left(2 z_\psi +  z_{_\Omega} \right)^3  ~~.
\nonumber 
\eear 
The second Dynkin index for the symmetric tensor is $T_2( \Omega )= 17/2$ \cite{Eichten:1982pn},
so the $[SU(15)_{\rm p} ]^2  U(1)$ anomaly vanishes  for  $z_{_\Omega}   = -19/17 z_\psi$.   
As a result, $A_{\rm preon} = A_{ \psi\psi\Omega} =  171 (15 z_\psi /17)^3$.
Likewise, the gravitational-$U(1)$ anomalies of the preons,  $285   z_\psi +   120  z_{_\Omega} $,
and of the prebaryons,  $171  (2 z_\psi +  z_{_\Omega}  )$, are equal.
This nontrivial matching makes it likely that the $SU(15)_{\rm p} $ dynamics indeed generates the 
171 $  \psi_i \psi_j \,  \Omega  $ chiral prebaryons. 
Additional evidence is provided by large-$N$ arguments \cite{Eichten:1985fs},
and by the complementarity between the Higgs and confining phases  \cite{Dimopoulos:1980hn}.

The preons shown in Table \ref{table:preons} lead to 
the $SU(3)_c \times SU(2)_W \times U(1)_Y$ charges of the prebaryons listed in Table \ref{table:quarks}.   
These composite quarks and leptons are written as left-handed fermions, using the notation 
$ \psi_U   \psi_i \, \Omega \equiv   \Omega_{Ui} \, $ for $i = 1, ..., 4$,  $ \psi_Q   \psi_U  \, \Omega \equiv   \Omega_{QU} $,  etc.
Some prebaryons have different charges even if they are  made of the same preons; 
for example,   $\Omega_{UD}  $ is a color triplet while another prebaryon with the same constituents, labelled $ \Omega_{UD}^\prime $, 
belongs to $\overline{6}$ of $SU(3)_c$.  Fermion anti-commutation 
prevents certain representations, {\it e.g.}, $ \Omega_{UU}  $ 
cannot be a $\overline{6}$. 

%%%%%%%%%%%%%%%%%%%%%
\begin{table}[t]
\begin{center}
\renewcommand{\arraystretch}{1.5}
\begin{tabular}{|c|c|c|}\hline    
 \ prebaryons \ &     $SU(3)_c \times SU(2)_W $   &  $U(1)_Y$    
\\ \hline \hline
       $ \Omega_{Q i} $  ,  $ \Omega_{Q E }  $   &  $(  3 , 2) $  &     \    $+1/6$  ,   $+7/6$ \         
\\ \hline
     $ \,  \Omega_{DL }  $   ,   $  \Omega_{U L } $       &     $( \overline 3 , 2) $ &     $-1/6$ ,    $-7/6$      
\\ \hline
    $ \Omega_{Ui }  $ ,    $  \, \Omega_{DE } $    &     $( \overline 3 , 1) $ &         $-2/3$ , $+4/3$
\\ \hline
     $ \, \Omega_{ DD}  $  ,  $  \Omega_{UU } $        &     $(    3 , 1) $ &      $+2/3$   ,       $-4/3$ 
\\ \hline
   $ \,  \Omega_{ Di }  $  ,  $ \Omega_{U E }  $    &     $(  \overline 3 , 1) $ &         $+1/3$ 
 \\ \hline 
    $  \Omega_{ UD}   +  \Omega_{UD }^\prime $      &     $(  3 , 1) + (  \overline 6 , 1) $ &     $-1/3$
  \\ \hline
       $  \Omega_{QQ }   +  \Omega_{QQ }^\prime  $    &  $ ( \overline 3 , 3) + (  6 , 1)  $  &    $+1/3$
 \\ \hline
       $  \Omega_{ QL }   +  \Omega_{QL }^\prime   $    &  $ (  3 , 1) + ( 3 , 3)  $  &     $-1/3$        
 \\ \hline
    $  \Omega_{QU }   +  \Omega_{QU }^\prime    $       &     $( 1 , 2) + (8 , 2)  $   &          $-1/2$ 
     \\ \hline
    $  \Omega_{QD }  + \Omega_{QD }^\prime     $       &     $( 1 , 2) + (8 , 2)  $   &           $+1/2$
 \\ \hline 
     $    \Omega_{ Li }    $  ,  $   \Omega_{LE }    $     &  $  ( 1, 2)  $  &     $-1/2$     ,   $+1/2$     
 \\ \hline
      \  $   \Omega_{Ei } \, ,    $      $     \Omega_{ ij}   \, ,     $        $  \Omega_{LL }    $   \    &  $  ( 1, 1)  $  &     $+1$   ,  0 ,   $-1$         
 \\ \hline
 \end{tabular}
% \vspace{0.1cm}
\caption{Quarks and leptons as prebaryons of $SU(15)_{\rm p}$. The preon flavor index $i=1,...,4$ implies four $\Omega_{Qi}  \equiv  \psi_Q   \psi_i \, \Omega    \, $
and six $\Omega_{ij}  \equiv  \psi_i  \psi_j \, \Omega    \, $ prebaryons. The prime on prebaryons such as $\Omega_{UD }^\prime $ denotes a higher representation.
Besides vectorlike pairs, the composite fermions form  3 SM generations. %\\ [-0.7cm] 
}
\label{table:quarks}
\end{center}
\end{table}
%%%%%%%%%%%

There are four prebaryons of the type  $ \Omega_{Qi}$, which transform  in the $ (  3 , 2, +1/6) $ representation of the SM gauge group, and one prebaryon, $\Omega_{DL} $, in the conjugate representation.
The latter forms a vectorlike pair with one linear combination of the four $\Omega_{Qi}$ prebaryons and acquires a mass, as discussed later on. Thus, there are 3 SM chiral quark doublets and 
one vectorlike quark (``Vquark") doublet of hypercharge 1/6. 

Similarly,  there are four $\Omega_{Ui}  $ prebaryons, which transform as $ ( \overline 3 , 1, -2/3) $, and one prebaryon, $\Omega_{DD} $,    in the conjugate representation. 
Hence, there is an up-type Vquark, and 3 chiral fermions identified as the SM up-type weak-singlet quarks. 
The down-type quark sector includes five prebaryons transforming as $( \overline 3 , 1, +1/3) $, 
and two prebaryons in the conjugate representation, forming two  Vquarks  and 3 SM quark  singlets.  

In the composite lepton sector, there are two vectorlike and 3 chiral weak doublets,
as well as one vectorlike and 3 chiral fermions transforming as $(1,1,+1)$. There are also 
6 SM singlet fermions, which may acquire both Dirac and Majorana masses.
All other composite quarks and leptons are in exotic vectorlike representations. 
Consequently, the $SU(15)_{\rm p}$  model 
implies the existence of three SM generations of composite quarks and leptons.

\medskip\medskip

%%%%%%%%%%%%%%%%%%%%%%%%%%%%%%%%%%%%%%%%%%%%%%%%%%%%%%%%%%%%%%
%%%%%%%%%%%%%%%%%%%%%%%%%%%%%%%%%%%%%%%%%%%%%%%%%%%%%%%%%%%%%
\section*{\large 3. \ Vectorlike  fermion  spectrum}
%\label{sec:VLQ}

The gauge theory presented in Table \ref{table:preons} has a low-energy behavior that includes composite fermions with 
the same quantum numbers as the SM quarks and leptons.
Its composite scalar sector is more difficult to analyze. Since 
the $SU(15)_{\rm p}$ dynamics probably preserves  the chiral symmetry of the preons,
it does not produce pion-like states. The only bosons 
that can be composed of two fields are vector ``premesons", 
such as $\overline \psi_i \gamma^\mu \psi_j$ or $\overline \Omega \gamma^\mu \Omega$.
The simplest spin-0 premesons are of the type  $\overline \psi_i \gamma^\mu T^a \psi_j D^\nu {\cal G}^a_{\mu\nu}$, 
where $ {\cal G}^a_{\mu\nu}$ is the $SU(15)_{\rm p}$ gauge field strength.
All these premesons are expected to have masses of order $\Lambda_{\rm pre}$.

Nonetheless, the low-energy theory may include composite scalars which are not premesons,  
but rather bound states of two prebaryons. These ``di-prebaryons" differ from  the 
deuteron in QCD in several ways. While the deuteron is a nonrelativistic 
nucleus bound mainly by long-range pion exchange, the di-prebaryons are relativistic states bound  by short-range remnants of the $SU(15)_{\rm p}$ dynamics. These remnant interactions can be thought of as vector premeson exchanges, although it is more accurate to 
view the di-prebaryons as 6-preon states bound by the $SU(15)_{\rm p}$ gauge field.
In the simplified picture of one premeson exchange, where the
premeson has vector couplings to the prebaryons, the most deeply bound di-prebaryons are scalars.

The di-prebaryons are more deeply bound in channels where the SM gauge interactions are attractive,
with the binding potential roughly given by 1-boson exchange \cite{Raby:1979my}:
\be
- \frac{1}{2 \, r} \left(C_3 \alpha_s+ C_2\alpha_2 +C_1 \alpha_Y \right)  ~~.
\ee
The $C_k$ coefficients are  determined by the SM charges of the two prebaryons, and 
$\alpha_s, \alpha_2,  \alpha_Y$ are the $SU(3)_c\times SU(2)_W\times U(1)_Y$ 
coupling constants at a scale of order $\Lambda_{\rm pre}$.
If the sum of bindings due to premesons and SM bosons 
is large enough, then the composite scalar acquires a negative 
squared mass and develops a VEV \cite{Bardeen:1989ds}.

Consider  the color-octet  prebaryons $ \Omega_{QU }^\prime $ and  $ \Omega_{QD }^\prime  $, which form a gauge-singlet scalar  $\phi_{88}$ with SM binding coefficients $\{ C_3, C_2, C_1 \} = \{ 6, 3/2, 1/2 \}  $.
The large QCD binding in this channel makes it likely that $\phi_{88}$ develops a VEV,
$\langle \phi_{88}  \rangle$.
Since premeson exchange 
interactions act at short distance and are strong but non-confining,  the effective low-energy theory
includes a large Yukawa coupling between  $\phi_{88}$ and its two constituents:  $\phi_{88}^\dagger  \Omega_{QU }^\prime \Omega_{QD }^\prime  $. 
Consequently, $\langle \phi_{88}  \rangle$
induces a Dirac mass for  $ \Omega_{QU }^\prime $ and  $ \Omega_{QD }^\prime  $,
which presumably is not much below $\Lambda_{\rm pre}$.
This large mass deters the formation of deeply bound scalars involving one color-octet  prebaryon and 
a prebaryon which is not an octet. 

Similarly,  the weak-triplet prebaryons form a scalar singlet $\phi_{33} \equiv  \Omega_{QQ } \Omega_{QL }^\prime$ 
with slightly weaker SM binding, $\{ C_3, C_2, C_1 \} = \{ 8/3, 4, 2/9 \}  $.
The color sextets form a scalar $\phi_{6\overline 6} \equiv  \Omega_{UD }^\prime  \Omega_{QQ }^\prime$ with $\{ C_3, C_2, C_1 \} = \{ 5, 0, 2/9 \}  $. 
As the three composite scalars discussed so far have the largest bindings and 
break the chiral symmetries of their prebaryon constituents,
the Vquarks of SM charges $(8,2, 1/2)$,
 $(3,3, -1/3)$ and  $(6,1, +1/3)$ are the heaviest composite fermions from Table \ref{table:quarks}.

 \begin{figure}[t!]
  \begin{center}
 \includegraphics[width=0.65\textwidth, angle=0]{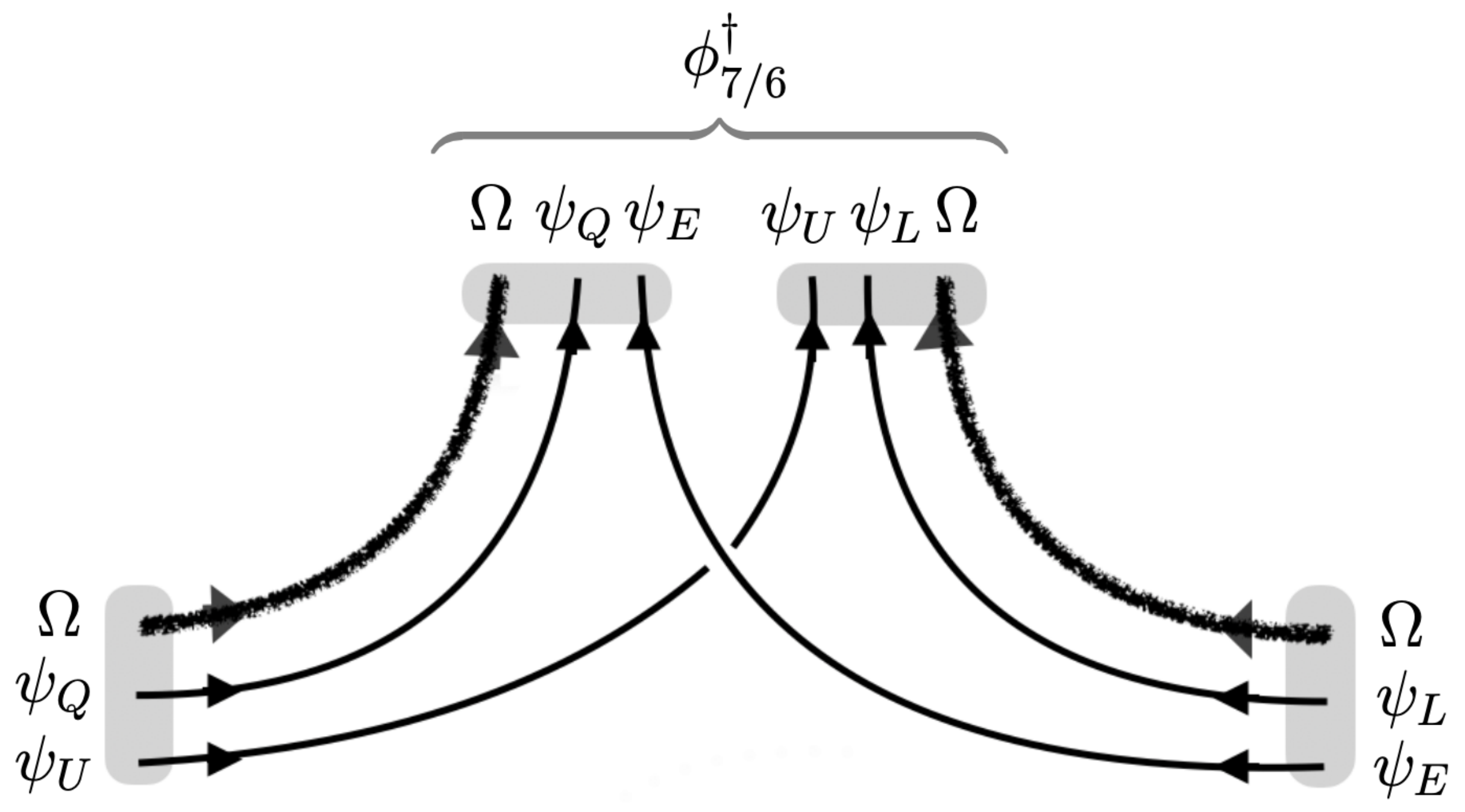}
 \vspace*{0.4cm}
   \caption{Effective Yukawa coupling $\phi_{7/6}^\dagger  \Omega_{QU }  \Omega_{LE}$, originating from a 4-prebaryon operator induced by $SU(15)_{\rm p}$ dynamics. }    %   \\ [-1cm] }
  \label{fig:Yukawa}
  \end{center}
\end{figure}

The large-$N$ expansion indicates that theories with strongly-coupled but nonconfining 
attractive interactions have a second-order chiral phase transition 
\cite{Bardeen:1993pj}, so the VEVs of composite scalars with 
weaker binding are suppressed or even 0. It is then reasonable to
assume that the only other vectorlike pairs that acquire masses
by coupling to the corresponding di-prebaryon 
are the ones that carry color and either are weak doublets or have the largest hypercharge.
The corresponding scalars are gauge singlets labelled by $\phi_{7/6}$, $\phi_{1/6}$ and 
$\phi_{4/3}$, where the  index refers to  the Vquark hypercharge.
These scalars have large Yukawa couplings to their constituents:
$\phi_{7/6}^\dagger  \Omega_{QE } \Omega_{UL }  $, $\phi_{1/6}^\dagger  \Omega_{Q4 } \Omega_{DL }  $,
$\phi_{4/3}^\dagger  \Omega_{DE } \Omega_{UU } $.
Here $ \Omega_{Q4 } $ represents the relabelled linear combination of the $ \Omega_{Qi}$'s that  
is the vectorlike partner of $\Omega_{DL } $.  Due to weaker SM binding,
these scalars have smaller VEVs than $\phi_{33}$  or  $\phi_{6\overline 6}$.
Hence, there are three  Vquarks of intermediate mass, of SM charges $(3,2, +7/6)$,
 $(3,2, +1/6)$,  $(3,1, -4/3)$.

The chiral symmetries of the prebaryons are sufficiently broken by the VEVs discussed so far  
such that all charged vectorlike pairs formed out of prebaryons listed in Table \ref{table:quarks} acquire masses.
To see that, consider first the color singlets $\Omega_{QU }$ and   $\Omega_{QD }$.
The interaction of the octets with $\phi_{88}$, upon exchanging a gluon, induces a smaller Yukawa  coupling
 $\phi_{88}^\dagger  \Omega_{QU } \Omega_{QD } $.
The color singlets $\Omega_{QU }  $ and $\Omega_{LE}$ have together the same 
preon content as $\phi_{7/6}$, and thus a Yukawa coupling $\phi_{7/6}^\dagger  \Omega_{QU }  \Omega_{LE}$
arises as in Fig.~\ref{fig:Yukawa}.  
 A $\phi_{1/6}^\dagger  \Omega_{QD }  \Omega_{L4}$ coupling is also induced, so that there are two 
vectorlike lepton (``Vlepton") doublets with a $2\times 2$ mass matrix. One mass eigenstate is mostly  $\Omega_{QU }\Omega_{QD }$,
with a mass probably comparable to that of the doublet Vquarks.
The other  mass eigenstate of charges $(1,2,-1/2)$ is much lighter, and  up to a small mixing is given by $ \Omega_{LE}\Omega_{L4}$.
The prebaryons of hypercharge $\pm 1/3$ form two Vquarks with a $2\times 2$ mass matrix,
whose entries come from the effective Yukawa  couplings
\be
\phi_{7/6}^\dagger  \Omega_{QL} \Omega_{UE}  \;  , \; \; \;   \phi_{1/6}^\dagger  \Omega_{D4 } \Omega_{QL }  \;  , \; \; \; 
\phi_{4/3}^\dagger  \Omega_{UD } \Omega_{UE }   ~~.
\label{eq:Vquark13}
\ee
 
At the scale $\Lambda_{\rm pre}$, 4-fermion operators are induced among any two prebaryon pairs made of the same preons ({\it e.g.}, see Fig.~\ref{fig:Yukawa}).
The operator $(\Omega_{U4}  \Omega_{DD})(\Omega_{UD}  \Omega_{D4})$  generates 
a 1-loop mass for $\Omega_{U4}  \Omega_{DD}$ proportional to the $\Omega_{UD}  \Omega_{D4}$ 
mass mixing, which arises from the terms (\ref{eq:Vquark13}). 
Another 4-fermion operator generates a 1-loop mass for $\Omega_{E4}  \Omega_{LL}$ proportional to the mass of the lightest
 Vlepton doublet, $\Omega_{LE}  \Omega_{L4}$.
As a result, the lightest Vquark  and charged Vlepton transform as $(3,1,+2/3)$ and $(1,1,+1)$.      

Current LHC searches \cite{CMS:2019eqb} cover  Vquarks 
of this type, but dedicated searches for  Vleptons remain to be performed.  
The  lightest  charged  Vlepton and Vquark  
may have masses of the order of 0.5 TeV and 2 TeV, respectively. The presence of several other 
composite vectorlike fermions with larger, hierarchical masses indicates that the compositeness scale 
 satisfies $\Lambda_{\rm pre} \gtrsim O(100)$ TeV.

\medskip\medskip
 
%%%%%%%%%%%%%%%%%%%%%%%%%%%%%%%%%%%%%%%%%%%%%%%%
%%%%%%%%%%%%%%%%%%%%%
\section*{\large 4. \ Composite Higgs sector}
%\label{sec:Higgs}

The $SU(15)_{\rm p}$ model analyzed so far has ingredients that may lead to a potentially realistic 
composite Higgs sector at low energy \cite{Dobrescu:1997nm}:  up-type  Vquarks
and strongly-coupled 4-fermion operators 
involving both vectorlike and SM quarks. There are multiple ways in which 
a composite Higgs sector could be generated, and given the uncertain behavior of
chiral gauge theories, it is hard to make rigorous statements. As an existence proof, 
a simple path towards a viable Higgs sector is sketched here.

The $SU(4)$ flavor symmetry of the $\psi_i$ preons is broken down to $SU(3)$ 
by  the VEV of $ \phi_{1/6} $, which is an $ \Omega_{Q4 } \Omega_{DL }$  %  
bound state.
The linear combinations of the $\Omega_{Qi}$ prebaryons which are not the vectorlike partner of 
$\Omega_{DL}$ are relabelled as $q_L^i$,  $ i = 1,2,3$, 
and identified as the SM left-handed quark doublets. 
Similarly, $u^{c \, i}$ are the linear combinations of $\Omega_{Ui}$ which remain  chiral,
and represent  the SM up-type quark singlets, taken as left-handed fermions.

To accommodate the known quark masses, the $SU(15)_{\rm p}$ model must include 
some interactions that break the $SU(3)$ flavor symmetry. An example is given by a scalar ${\cal A} $ 
that transforms under  $SU(15)_{\rm p}$ as the conjugate of the antisymmetric 2-tensor, {\it i.e.}, in the $\overline{105}$ representation. 
Even for a mass $M_{\cal A} <  \Lambda_{\rm pre}$, the scalar 
does not affect the confining dynamics of $SU(15)_{\rm p}$, because the $  {\cal A} \, \psi_i \psi_j $ premesons  
are not lighter than $\Lambda_{\rm pre}$, and  the spectrum of chiral prebaryons is not modified. 

The most general Yukawa couplings of ${\cal A} $, up to a flavor transformation of $\psi_i$, $i=1,2,3$,
can be written as
\be
 {\cal A} \, \psi_4  \left( \lambda_{44}    \psi_4 + \lambda_{43}  \psi_3 \right) +  {\cal A}  \sum_{i\ge j=1}^{3} \lambda_{ij} \psi_i  \psi_j ~~,
\ee
where $\lambda_{ij} $ are dimensionless parameters.
Integrating out ${\cal A} $  produces 4-preon operators, such as 
$(\psi_4  \psi_3) (\overline \psi_4 \overline \psi_3)$,
which upon  a Fierz transformation and to leading order in $1/N$ (here $N = 15$) becomes \cite{Hill:1991at}
\be
- \frac{ | \lambda_{43} |^2}{M_{\cal A} ^2}  \left( \overline  \psi_3 \sigma^\mu T^a \psi_3    \right)   \left( \overline  \psi_4 \sigma_\mu T^a \psi_4   \right) ~.
\label{eq:34operator}
\ee
This attractive interaction bolsters the formation of di-prebaryons, especially  if $|\lambda_{43}|  \gtrsim  O(1)$ and 
$M_{\cal A} \lesssim \Lambda_{\rm pre}$. 

The $\Omega_{U4}$ component of the lightest  Vquark may form 
a di-prebaryon  with $\Omega_{Q3}$, due to premeson exchange plus SM binding, 
which has coefficients $\{8/3,0,2/9 \}$. The additional attraction provided by 
${\cal A} $ exchange may be sufficient for  the squared mass of this
 scalar to turn negative.  The SM charges of the  $\Omega_{U4}\Omega_{Q3}$ 
 scalar are $(1,2,-1/2)$, so it is appropriate to call it the up-type Higgs doublet, $H_u$.
Several other scalars with the same quantum numbers may form as 
$\Omega_{Ui}\Omega_{Qj}$ bound states, but their squared masses may be positive and large  if 
$|\lambda_{ij}|^2 \ll |\lambda_{43}|^2$.
The $H_u$ doublet has a large Yukawa coupling to $q_L^3$ and $ \Omega_{U4} $. In addition, ${\cal A} $ exchange  leads to
 effective Yukawa couplings of the type  $H_u^\dagger \,  u^{c i}  q_L^j $, with coefficients proportional to     
 $\lambda_{ij} \lambda_{43}^\dagger   $.
Therefore, a mass matrix for the SM up-type quarks is generated.
The observed charm to top quark mass ratio requires $|\lambda_{22}| \ll |\lambda_{33} |$.

Operator (\ref{eq:34operator}) together with premeson and SM boson exchanges also 
produces a  scalar $H_d \equiv \Omega_{D4} \Omega_{Q3}$, which
induces a mass matrix for the SM down-type quarks. Given that the Vquarks of charge $-1/3$
are heavier than the Vquark of charge 2/3,
the $H_d$ VEV is suppressed compared to the $H_u$ VEV. 
Subleading contributions to the fermion mass matrices arise from operators involving $H_d H_u^\dagger H_u$. 
Those may sufficiently contribute to the strange quark mass to account for the 
relation between observed quark mass ratios at the TeV scale: $m_s/m_b \approx 5 m_c/m_t$ \cite{Xing:2011aa}.
The subleading contributions also lead to a slight misalignment of the up- and down-type quark mass matrices,
resulting in off-diagonal elements of the CKM matrix.

Four-preon operators mediated  by the $ {\cal A}$ scalar, such as 
$\lambda_{11}^\dagger  \lambda_{22} / M_{\cal A}^2 (\psi_1 \psi_1) (\overline \psi_2 \overline \psi_2)$, 
lead to 4-quark operators that contribute to flavor-changing processes. Experimental constraints 
on the 4-quark operators are satisfied given that $M_{\cal A} \sim \Lambda_{\rm pre} \gtrsim O(100)$ TeV and 
$\lambda_{11} \ll \lambda_{22}  \ll  1$. 

The SM lepton masses arise from effective Higgs Yukawa couplings, 
whose origin is more intricate. The dimension-8 operator  $\phi_{7/6} \phi_{88}^\dagger \left( \psi_D  \psi_Q \right)\left( \overline\psi_L  \overline\psi_E \right) $ is generated by $SU(15)_{\rm p}$ dynamics at the $\Lambda_{\rm pre}$ scale. 
At lower scales, this operator produces the 4-prebaryon operators  
$ (\overline\Omega_{D4}  \overline \Omega_{Q3}) \left( \Omega_{L4} \Omega_{E3} \! +  \Omega_{L3} \Omega_{E4}  \right)$,
$ ( \overline\Omega_{D3}  \overline \Omega_{Q3} )( \Omega_{L3} \Omega_{E3} )$
and $ ( \overline\Omega_{D4}  \overline \Omega_{Q4} )( \Omega_{L4} \Omega_{E4})$,
which via ${\cal A}$ exchange lead to the $\tau$ Yukawa  coupling, $H_d^\dagger \,  e^{c3} \ell_L^3$. Therefore, $m_\tau$ is suppressed by $\langle \phi_{7/6} \rangle \langle \phi_{88} \rangle /\Lambda_{\rm pre}^2$ 
compared to $m_b$, but this is partially compensated by the multiplicity of contributions, so $m_\tau/m_b$ may be large enough. 

The mechanisms for fermion mass generation sketched above rely on the existence of two Higgs doublets.
Other composite scalar doublets, such as $\Omega_{L4} \Omega_{E3}$, may also form;
even if these do not have VEVs, they might mediate additional effective 
couplings of the fermions to $H_u$ and $H_d$.
As the chiral phase transition is of second order, it is difficult to estimate the composite scalar masses. 

The attractive interaction (\ref{eq:34operator}) and the analogous one proportional to $| \lambda_{44} |^2$ 
are amplified by a factor of 3 or 4 
in the self-interactions of $\Omega_{43}$ and $\Omega_{44}$, leading to gauge-singlet scalars with negative squared masses.
Their VEVs then induce Majorana masses for $\Omega_{43}$ and $\Omega_{44}$, which may be near $\Lambda_{\rm pre}$.
The same VEVs, after an ${\cal A}$ exchange, give masses to the 
other four $\Omega_{ij}$ prebaryons. These composite gauge-singlet fermions 
participate in a seesaw mechanism responsible for the neutrino masses.

\medskip\medskip\medskip

%%%%%%%%%%%%%%%%%%%%%%%%%%%%%%%%%%%%%%%%%%%%%%%%%%%%%%%%%%%%%%%%%%
%%%%%%%%%%%%%%%%%%%%%%%%%%%%%%%%%%%%%%%%%%%%%%%%%%%%%%%%%%%%%%%%%%%
\section*{\large 5. \ $SO(10)$ unification}

The preon fields include 60 color triplets,  so  the QCD coupling is not asymptotically free above $\Lambda_{\rm pre}$.
Even below that, the sextet and octet quarks turn the  $SU(3)_c$ $\beta$ function positive,
indicating that the SM group must be embedded in a larger gauge group. It turns out 
that the preon field content from Table \ref{table:preons} allows $SO(10)$ unification. Moreover, the SM gauge 
couplings evolve roughly towards a common value at a scale $\Lambda_{10} > \Lambda_{\rm pre}$, albeit 
establishing coupling unification is difficult
due to nonperturbative effects. 

The  field content of the theory above $\Lambda_{10}$ is the following: 
$SU(15)_{\rm p}  \times  SO(10)$ gauge group, with fermions 
\be
\Psi \in  (15,16)   \; ,   \;\;  \;\; \psi_2,\psi_3,\psi_4 \in (15,1)      
\; ,  \;\;  \;\; \Omega \in (\overline{120},1)  ~,
\label{eq:10preons}
\ee
a scalar ${\cal A} \in (\overline{105},1)$, and two scalars whose VEVs break $SO(10)$ 
and transform, {\it e.g.}, as $(1,16)$ and  $(1,45)$.  
Both $SU(15)_{\rm p}$ and $SO(10)$ are asymptotically free. 
The prebaryons listed in Table \ref{table:quarks} form an $SO(10)$ representation:
$3\times (16 +1) + 120$, where 120 contains all $\Psi\Psi \Omega$ states.

The SM singlet component of $\Psi$ is $\psi_1$. Consequently, $\lambda_{i1}= 0 $ due to the $SO(10)$ symmetry,
which implies that the first generation fermions are naturally lighter than the second generation ones.
The origin of the first generation masses may be a dimension-5 coupling of ${\cal A}  \, \psi_i  \Psi  $ to the $(1,16)$ scalar,
which arises from renormalizable couplings of an additional scalar or fermion.

Interestingly, proton decay operators are not generated at the unification scale.
The dimension-6 operator $qqq\ell$ is a 4-prebaryon operator 
$ \left( \Omega_{Qi} \right)^3 \! \Omega_{Li}$, which could arise from 
the dimension-18 operator $\left( \psi_Q \psi_i  \Omega \right)^3 \!\psi_L \psi_i \Omega$. The latter is not mediated 
by $SO(10)$ gauge bosons. 
Same argument applies to the  $u^c u^c d^c e^c$ operator. Thus,
$\Lambda_{10}$ can be substantially below the usual \cite{Mohapatra:1986uf} unification scale.      

To conclude, the $SU(15)_{\rm p} \times SO(10)$ gauge theory with the preons shown
in (\ref{eq:10preons}) produces light prebaryons transforming as 3 SM generations of fermions. 
Several composite vectorlike fermions have a hierarchical mass spectrum.
A potentially viable Higgs sector likely arises from bound states of two prebaryons.   
Further studies of this theory of composite quarks and leptons are warranted.
Particularly useful would be an improved
understanding of strongly-coupled chiral gauge theories ({\it e.g.}, see \cite{Grabowska:2016bis, Bolognesi:2021jzs}),
and experimental searches for the composite scalars and vectorlike fermions predicted here.

\bigskip\bigskip\bigskip\bigskip

%%%%%%%%%%%%%%
\noindent
{\it  Acknowledgments:}  {\small  I would like to thank Benoit Assi, Yang Bai, Sekhar Chivukula, Patrick Fox and Kyoungchul Kong for insightful comments. 
This work is supported by Fermi Research Alliance, LLC under Contract DE-AC02-07CH11359 with the U.S. Department of Energy.
}

%\bigskip

%%%%%%%%%%%%%%%%%%%%%%%%%%%%%%%%%%%%%%%%%%  
\providecommand{\href}[2]{#2}\begingroup%\raggedright

\vfil
%%%%%%%%%%%%%%%%%%%%%%%%%%
\end{document}